\documentclass[preprint,showpacs,preprintnumbers,amsmath,amssymb]{revtex4}

\usepackage{graphicx}
\usepackage{dcolumn}
\usepackage{bm}

\begin{document}

\preprint{APS/123-QED}

\title{Why Does Undoped FeSe Become  A High $T_c$ Superconductor Under Pressure?}

\author{T. Imai$^{1,2}$, K. Ahilan$^{1}$, F. L. Ning$^{1}$,  T. M. McQueen$^{3}$, and R. J. Cava$^{3}$}

\affiliation{$^{1}$Department of Physics and Astronomy, McMaster University, Hamilton, Ontario L8S4M1, Canada}
\affiliation{$^{2}$Canadian Institute for Advanced Research, Toronto, Ontario M5G1Z8, Canada}
\affiliation{$^{3}$Department of Chemistry, Princeton University, NJ 08544, USA}

\date{\today}

\begin{abstract}
Unlike the parent phases of the iron-arsenide high $T_c$ superconductors, undoped FeSe is not magnetically ordered and exhibits superconductivity with $T_{c}\sim9$~K.  Equally surprising is the fact that applied pressure dramatically enhances the modest $T_{c}$ to $\sim 37$~K.  We investigate the electronic properties of FeSe using $^{77}$Se NMR to search for the key to the superconducting mechanism.  We demonstrate that the electronic properties of FeSe are very similar to those of electron-doped FeAs superconductors, and that antiferromagnetic spin fluctuations are strongly enhanced near $T_c$.  Furthermore, applied pressure enhances spin fluctuations.  Our findings suggest a link between spin fluctuations and the superconducting mechanism in FeSe.\end{abstract}


\maketitle


The discovery of new iron-arsenide (FeAs) high $T_c$ superconductors \cite{Kamihara} has led to a frenzy of research over the last year \cite{Norman}.  The superconducting mechanism still remains enigmatic, but it has become clear that all FeAs superconductors share a salient feature; their undoped parent phase is magnetically ordered in a Spin Density Wave (SDW) state, and the superconducting phase emerges when electron or hole doping suppresses the SDW instability \cite{Norman}.  For example, both undoped LaFeAsO and BaFe$_2$As$_2$ have a SDW ground state below $\lesssim140$~K \cite{delaCruz,Rotter}, and the superconducting state in LaFeAsO$_{1-x}$F$_{x}$ ($T_{c}\sim 28$~K) \cite{Kamihara} and BaFe$_{2-x}$Co$_{x}$As$_{2}$ ($T_{c}\sim 22$~K) \cite{Sefat} requires $5 \sim 8$~\% of electron-doping.  Therefore one could speculate that residual spin fluctuations may be playing a key role in the superconducting mechanism.  Alternatively, one could also argue that magnetism and superconductivity are competing against each other.

In view of the possible link or competition between magnetism and superconductivity in FeAs high $T_c$ superconductors, superconductivity in FeSe ($T_{c}\sim 9$~K) \cite{Hsu, McQueen} raises interesting questions, and provides important test ground for the ideas to account for high $T_c$ superconductivity in iron-based systems \cite{Johannes}.  We note that the initial discovery identified $\alpha$-FeSe$_{1-\delta}$ with large deficiency $\delta \sim 0.12$ as the superconducting phase \cite{Hsu}, which led to a misperception that electron doping by the Se deficiency destroys a SDW ground state and stabilizes superconductivity.  However, as some of us have more recently shown, the apparently large $\delta$ is caused by oxygen contamination of the Fe ingredient \cite{McQueen}.  The actual superconducting phase is the stoichiometric $\beta$-Fe$_{1.01\pm0.02}$Se, or equivalently, $\beta$-FeSe$_{0.99\pm0.02}$ \cite{McQueen}, {\it i.e.} superconductivity in FeSe does not require electron doping.  Furthermore, application of pressure on FeSe raises $T_c$ to as high as $\sim$37~K \cite{Mizuguchi,Medvedev, Margadonna}.  These observations are counterintuitive if we compare the number of electrons at As and Se sites.  The nominal ionic state of the FeAs layers is [FeAs]$^{-}$ in the undoped parent phase with an SDW ground state ({\it e.g.} LaFeAsO and BaFe$_2$As$_2$), and the As$^{3-}$ sites have eight electrons in the (4s)$^2$(4p)$^6$ orbitals.  Since a Se atom has one extra electron compared to an As atom, we also expect that  eight electrons fill the (4s)$^2$(4p)$^6$ orbitals at Se$^{2-}$ sites in the stoichiometric FeSe.  This simple electron counting suggests that FeSe should also undergo a SDW rather than superconducting transition if analogies hold between FeAs and FeSe systems.  In fact, band calculations suggest that the Fermi surface nesting induces a SDW ground state in undoped FeSe \cite{Subedi}.   How different is the stoichiometric FeSe superconductor from electron or hole doped FeAs superconductors?  Is $T_c$ as low as $\sim 9$~K because spin fluctuations associated with the SDW instability are absent?  What is the driving mechanism behind the large enhancement of $T_c$ in FeSe under pressure?

In this {\it Letter}, we report a $^{77}$Se NMR investigation of FeSe.  We demonstrate that the electronic properties of the {\it undoped} FeSe ($T_{c}\sim 9$~K) share remarkable similarities with  {\it electron-doped} FeAs superconductors.  Our measurements of the $^{77}$Se spin-lattice relaxation rate, $1/T_1$, in ambient pressure indeed provide evidence for strong enhancement towards $T_c$ of antiferromagnetic spin fluctuations at finite wave vector ${\bf q}\neq {\bf 0}$. This finding suggests that undoped FeSe superconductor is actually on the verge of an SDW ordering.  Furthermore, we show that application of hydrostatic pressure {\it enhances} spin fluctuations as well as $T_c$.  These results strongly suggest that spin fluctuations have a strong link with the superconducting mechanism of FeSe. 

\begin{figure}
\includegraphics[width=12cm]{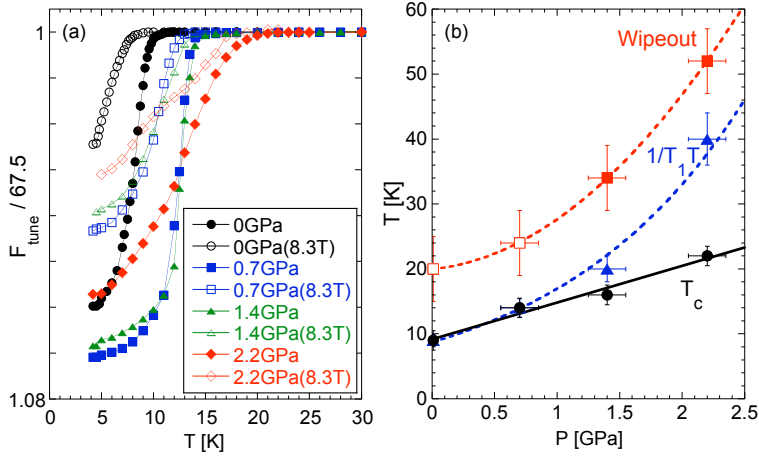}
\caption{\label{fig1:epsart}(Color Online) (a) AC susceptibility of $\beta$-Fe$_{1.01}$Se in $B_{ext}=0$ and 8.3~Tesla under various pressures.  We measured the increase of the tuning frequency $F_{tune}\sim 1/\sqrt{L(1+4\pi \chi') C}$ of the LC tank circuit used for NMR.   $F_{tune}$ is normalized by the normal state value, $F_{tune}(T>T_{c})=67.5$~MHz.  Superconducting diamagnetic susceptibility, $\chi' < 0$, enhances $F_{tune}$.  Notice that the vertical axis is reversed.   (b) The $T-P$ phase diagram of $\beta$-Fe$_{1.01}$Se.  Circles, triangles and squares represent $T_c$ in $B_{ext}=0$ (from Fig.1a), the hump of $1/T_{1}T$ (from Fig.4), and the onset of the loss of the NMR signal intensity ("wipeout", from Fig.2d), respectively.  All lines are guides for the eyes.}
\end{figure}

Our NMR sample is $\beta$-Fe$_{1+\delta}$Se with a nearly defect free composition of $\delta=0.01\pm0.02$ \cite{McQueen}.  For comparison, we also investigated a non-superconducting sample with $\delta=0.03$.   From the transport, specific heat, and SQUID measurements, $T_{c} \sim 9$~K for $\beta$-Fe$_{1.01}$Se, while the upper bound of $T_{c}$ is 0.4~K for $\beta$-Fe$_{1.03}$Se \cite{McQueen}.  Detailed structural studies based on x-ray and neutron diffraction measurements revealed no hint of impurity phases.  While conducting NMR measurements at each pressure, we also carried out AC susceptibility measurements at 67.5~MHz using the NMR coil within the high pressure cell.  As shown in Fig.1a and 1b, $T_c$ in zero applied magnetic field ($B_{ext}=0$) rises roughly linearly from $\sim 9$~K in ambient pressure ($P=0$~GPa) to $\sim 14$~K (0.7~GPa), $\sim 16$~K (1.4~GPa),  and $\sim 22$~K (2.2~GPa).  We carried out most of the NMR measurements in $B_{ext}=8.3$ or 9~Tesla, which has very little effect on $T_c$ except in ambient pressure.  We carried out  $1/T_{1}$ measurements by saturating the whole NMR line with comb pulses.  We found that the recovery of nuclear magnetization can be fitted with single exponential very well.  In $P=0$~GPa, we used $B_{ext}=1.5$~Tesla for $1/T_{1}T$ measurements below $T_c$  to minimize the suppression of $T_c$ by $B_{ext}$; the results above $T_c$ showed no dependence on $B_{ext}$.     

In Fig.2, we present representative powder-averaged $^{77}$Se NMR lineshapes for polycrystalline samples.  $^{77}$Se has nuclear spin $I=1/2$ with the nuclear gyromagnetic ratio $\gamma_{n}/2\pi = 8.118$~MHz/Tesla, hence it is expected to gives rise to a single NMR peak at the Zeeman frequency of $f_{o} \sim (\gamma_{n}/2\pi)B_{ext} \sim 67.4$~MHz in $B_{ext}=8.3$~Tesla, or $f_{o} \sim 73.0$~MHz in 9~Tesla.  The observed $^{77}$Se NMR linewidth of $\sim 0.03$~MHz in $\beta$-Fe$_{1.01}$Se is by a factor of $\sim3$ narrower than the earlier report for a highly disordered "FeSe$_{0.92}$" sample \cite{Tou}, and shows very little temperature dependence.  We also observed no distribution of $1/T_{1}$ in the normal state unlike the case of "FeSe$_{0.92}$".   These results assure us that our $\beta$-Fe$_{1.01}$Se sample is homogeneous and nearly defect free.  On the other hand, the NMR lineshape of $\beta$-Fe$_{1.03}$Se is somewhat broader, and becomes more broad at low temperatures without changing the integrated intensity.  This hints at the presence of defects, which may contribute to the suppression of $T_c$.

Our results in Fig.2 show that the actual NMR peak frequency, $f$, is shifted from $f_{o}$.  The shift, $\Delta f = f - f_{o}$, is temperature dependent.    We plot the temperature dependence of  the {\it Knight shift} $K = \Delta f/f_{o}$ in Fig.3.  The Knight shift arises because $B_{ext}$ polarizes the spin and orbital angular momenta of electrons in proportion to their magnetic susceptibilities, and these induced polarizations exert additional hyperfine magnetic fields on $^{77}$Se nuclear spins.  Generally, we can express $K = K_{spin}+K_{chem}$.  The spin contribution $K_{spin} = A_{hf}\chi_{spin}$ is proportional to the spin susceptibility, $\chi_{spin}$, in the FeSe layers ($A_{hf}$ is the hyperfine interaction between electrons and the $^{77}$Se nuclear spin).  The chemical shift $K_{chem}$ is generally temperature independent and caused by polarized orbital moments.  Thus our results in Fig.3 establish that $\chi_{spin}$ of the superconducting FeSe decreases almost linearly with temperature from 480~K to $\sim 100$~K, and then levels off.  The observed behavior of  $\chi_{spin}$ is similar to that of the electron doped LaFeAsO$_{1-x}$F$_{x}$ and Ba[Fe$_{1-x}$Co$_{x}$]$_{2}$As$_{2}$ superconductors \cite{Ahilan,NJP,Ning1,Ning2}.   In particular, our new results resemble the  $^{75}$As Knight shift in the optimally electron-doped superconductor Ba[Fe$_{0.92}$Co$_{0.08}$]$_{2}$As$_2$ \cite{Ning1,Ning2} ($T_{c}=22$~K, $K_{chem}=0.2 \sim 0.25 \%$ \cite{Ning1} and $A_{hf}\sim 20$~kOe/$\mu_{B}$ \cite{Kitagawa}).

\begin{figure}
\includegraphics[width=12cm]{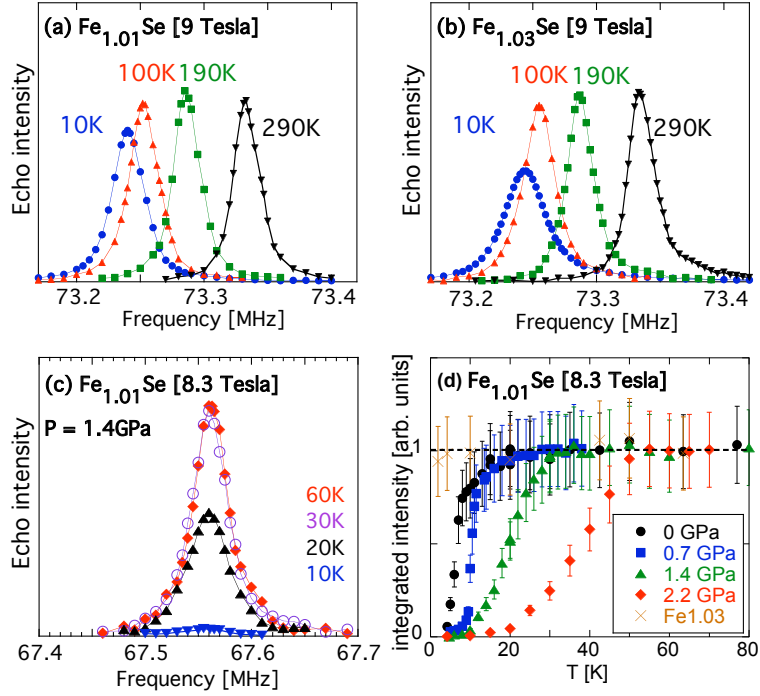}
\caption{\label{fig2:epsart} 
(Color Online)  Representative $^{77}$Se NMR lineshapes measured for (a) $\beta$-Fe$_{1.01}$Se (0~GPa), (b) $\beta$-Fe$_{1.03}$Se (0~GPa), and (c) $\beta$-Fe$_{1.01}$Se ($P=1.4$~GPa).  The intensity is corrected for  the Boltzman factor.  (d) The temperature dependence of the integrated NMR intensity of $\beta$-Fe$_{1.01}$Se.  For comparison, we also show the results for $\beta$-Fe$_{1.03}$Se (x). 
}
\end{figure}

How does $\chi_{spin}$ vary with applied pressure $P$ and the concentration $x$?  Our results in Fig.3 show that $K$, hence $\chi_{spin}$, changes little between $P=0$~GPa and 2.2~GPa.  Moreover, the non-superconducting $\beta$-Fe$_{1.03}$Se also exhibits nearly identical $\chi_{spin}$.  The inevitable conclusion from these findings is that the physical parameters that control $\chi_{spin}$ ({\it e.g.} the density of electronic states, Fe spin-spin exchange interaction $J$, {\it etc.}) may not have a direct link with the superconducting mechanism.  Generally, as is well known for high $T_c$ cuprate superconductors \cite{Johnston}, the growth of antiferromagnetic short-range order could suppress $\chi_{spin}$ with decreasing temperature.  In view of the absence of strong $P$ and $x$ dependencies of  $\chi_{spin}$ in Fig.3, it is tempting to conclude that such antiferromagnetic correlations, possibly caused by the nesting of Fermi surfaces \cite{Subedi}, may be irrelevant to superconductivity.  However, note that $\chi_{spin}$ is only a measure of the {\it uniform} ${\bf q}={\bf 0}$ response by electron spins to a uniform perturbation $B_{ext}$.  Furthermore, complicated Fermi surface geometry is likely to lead to coexistence of various ${\bf q}$ modes of spin excitations in the iron-based superconductors \cite{Scalapino}, hence $\chi_{spin}$ is not necessarily the best probe of magnetic correlations.   To explore the potential link between magnetism and the superconducting mechanism, one needs to measure the magnetic response of  the non-zero wave-vector modes, ${\bf q}\neq{\bf 0}$. 

\begin{figure}
\includegraphics[width=12cm]{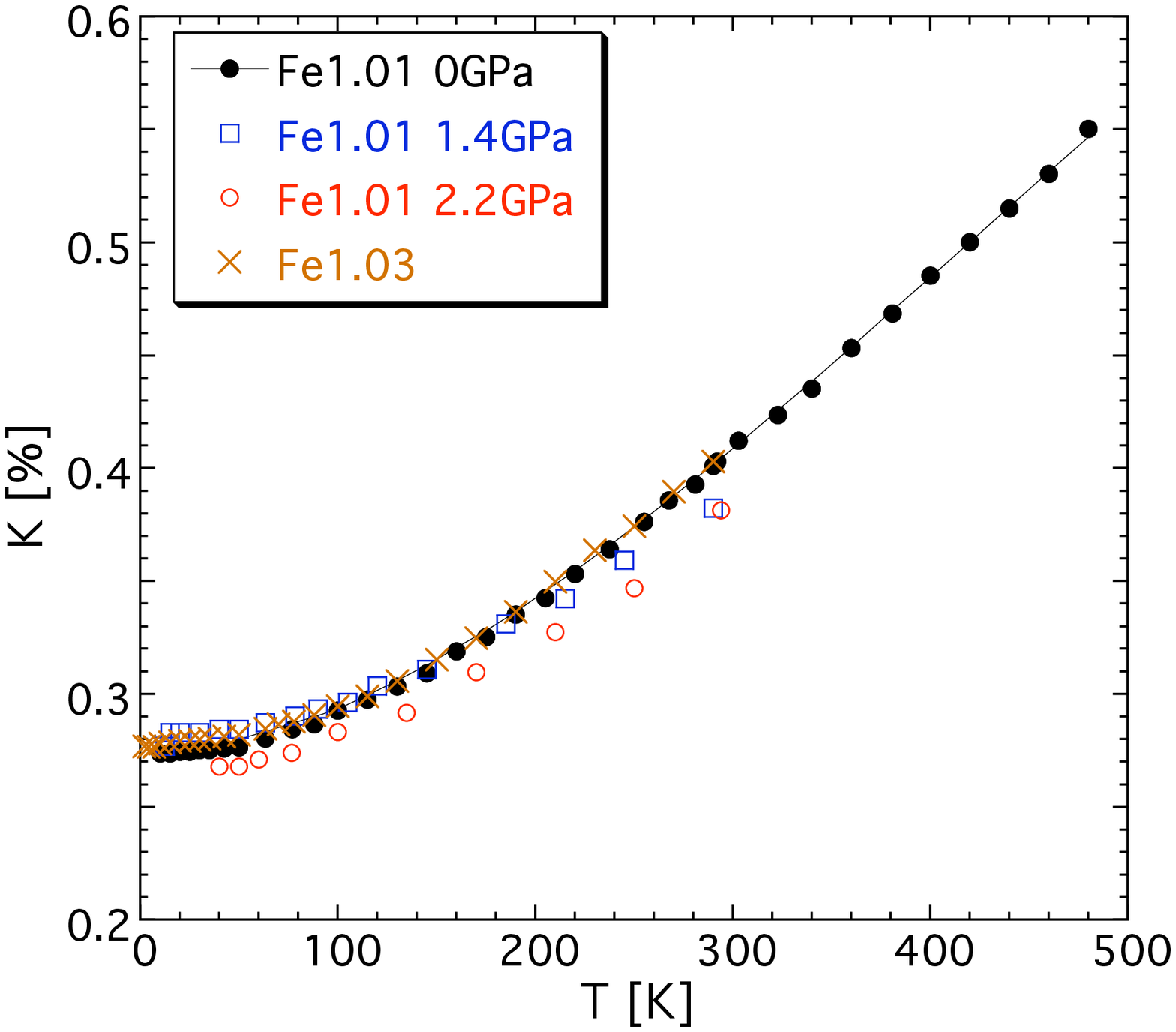}
\caption{\label{fig3:epsart} 
(Color Online) The temperature, pressure, and concentration dependencies of $^{77}$Se NMR Knight shift $K = K_{spin}+K_{chem}$ in the normal state above $T_c$.  $K_{spin}$ is proportional to the uniform spin susceptibility $\chi_{spin}$, and $K_{chem}$ is constant, hence the results reflect the temperature dependence of $\chi_{spin}$. \\
}
\end{figure}

 In Fig.4, we present the temperature dependence of $1/T_{1}T \propto \sum_{{\bf q}}|A_{hf}({\bf q})|^{2} \chi"({\bf q}, f)$, the nuclear spin-lattice relaxation rate $1/T_1$ divided by temperature $T$.   $A_{hf}(\bf q)$ and $\chi"({\bf q}, f)$ represent the wave vector ${\bf q}$-dependent hyperfine form factor \cite{Shastry} and the imaginary part of the dynamical electron spin susceptibility at the NMR frequency $f \sim 67.5$~MHz, respectively.  Thus {\it $1/T_{1}T$ measures the weighted average for various ${\bf q}$-modes of the low frequency spin fluctuations}.  $1/T_{1}T$ observed for superconducting $\beta$-Fe$_{1.01}$Se is strikingly similar to that of the optimally electron-doped Ba[Fe$_{1.92}$Co$_{0.08}$]$_{2}$As$_2$ \cite{Ning2}; $1/T_{1}T$ decreases with $T$ down to $\sim 100$~K, then begins to increase toward $T_{c}$.  Since $K$ is nearly temperature independent below 100~K, the latter implies that {\it some ${\bf q}\neq{\bf 0}$ antiferromagnetic modes of spin fluctuations are strongly enhanced toward $T_c$}.
 
 Recalling that removal of a few percent of electrons transforms the superconducting ground state of Ba[Fe$_{0.92}$Co$_{0.08}$]$_{2}$As$_2$ into a SDW ordered state \cite{Ning2}, the similarities of $K$ and $1/T_{1}T$ between $\beta$-Fe$_{1.01}$Se and Ba[Fe$_{0.92}$Co$_{0.08}$]$_{2}$As$_2$ lead us to conclude that {\it  superconductng FeSe is also in close proximity to a magnetic instability}.   We also note that $1/T_{1}T$ measured in 0 and 0.7~GPa shows a sharp peak exactly at $T_{c}(B_{ext})$ as determined by the AC susceptibility data presented in Fig.1a.  This means that {\it superconductivity sets in at $T_c$ after antiferromagnetic spin fluctuations are enhanced}, and {\it the opening of the superconducting energy gap suddenly suppresses low frequency spin fluctuations}.  Unlike typical isotropic BCS s-wave superconductors with a full gap, $1/T_{1}T$ measured in 0 and 0.7~GPa does not exhibit a Hebel-Slichter coherence peak just below $T_c$.  Instead, as shown in the inset to Fig.4, $1/T_{1}$ dives below $T_c$, exhibiting a power-law-like behavior.  This finding is consistent with earlier NMR reports on various iron-based superconductors \cite{Nakai,Ning1,Tou,NJP}.  

\begin{figure}
\includegraphics[width=12cm]{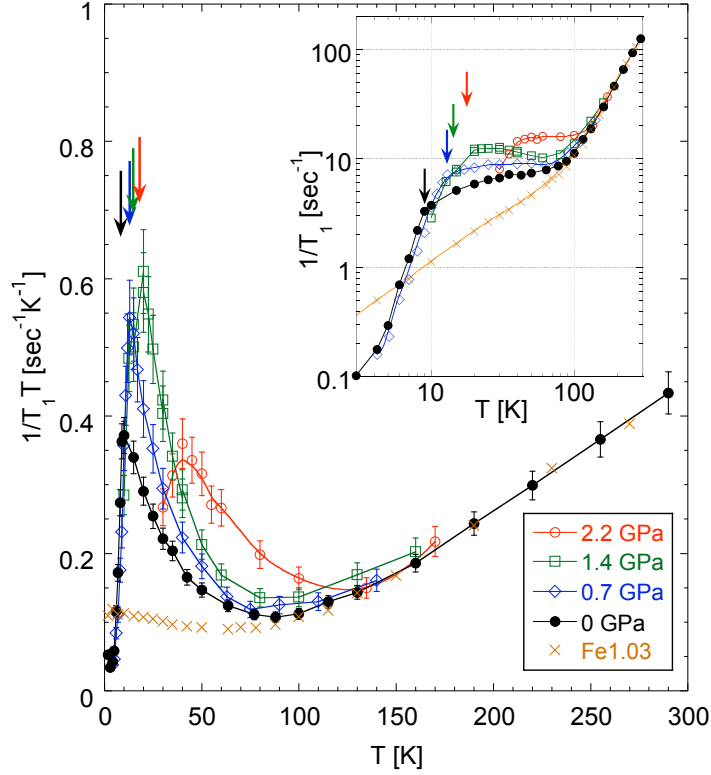}
\caption{\label{fig4:epsart} (Color Online)  $1/T_{1}T$ for superconducting $\beta$-Fe$_{1.01}$Se under various pressures, and for non-superconducting  $\beta$-Fe$_{1.03}$Se in $P=0$~GPa.  $1/T_{1}T$ reflects the spin fluctuation susceptibility averaged over various wave vector modes ${\bf q}$.  Inset : A log-log plot of $1/T_{1}$.  Vertical arrows mark (from left to right) $T_{c}$ for 0, 0.7, 1.4, and 2.2~GPa in the applied magnetic field.
}
\end{figure}

A sticky but essential question to address is whether our conclusions in the last two paragraphs imply that (a) antiferromagnetic spin fluctuations are positively linked with the superconducting mechanism, or (b) antiferromagnetic spin fluctuations are competing with superconductivity.   Two pieces of evidence seem to favor scenario (a).  First, $1/T_{1}T$ of $\beta$-Fe$_{1.03}$Se increases very little below 100~K, {\it i.e.} the enhancement of antiferromagnetic spin fluctuations below 100K in $\beta$-Fe$_{1.03}$Se, if any, is much weaker than in $\beta$-Fe$_{1.01}$Se.    Second, we find that both $T_c$ and spin fluctuations grow under pressure.  For example, $1/T_{1}T$ at 50~K increases from 0.15~sec$^{-1}$K$^{-1}$ in 0~GPa to 0.18~sec$^{-1}$K$^{-1}$ (0.7~GPa), 0.21~sec$^{-1}$K$^{-1}$ (1.4~GPa), and 0.31~sec$^{-1}$K$^{-1}$ (2.2~GPa).  If spin fluctuations with ${\bf q}\neq{\bf 0}$ are genuinely competing against the superconducting mechanism, we would expect a suppression of $1/T_{1}T$ under pressure when $T_c$ rises.  
 
Close inspection of the $1/T_{1}T$ data reveals that $1/T_{1}T$ shows a broad hump at $\sim 20$~K in 1.4~GPa and $\sim 40$~K in 2.2~GPa.  These humps are significantly above $T_{c}(B_{ext})=14.5$~K and 18~K as determined by AC susceptibility measured in identical conditions of $B_{ext}$ and $P$, hence we can't attribute the suppression of $1/T_{1}T$ below these humps to the opening of a superconducting energy gap.  Furthermore, we found that the integrated intensity of the NMR signal begins to decrease at temperatures somewhat above these humps (at $\sim 34$~K in 1.4~GPa and $\sim 50$~K in 2.2~GPa) as shown in Fig.2c and 2d.  We summarize the T-P phase diagram of $\beta$-Fe$_{1.01}$Se in Fig.1b including these anomalies.   Notice that the NMR signal intensity in 1.4 and 2.2~GPa is almost completely wiped out by the time we reach $T_c$.  Thus we need to be somewhat cautious in interpreting the  $1/T_{1}T$ results below these humps, because $1/T_{1}T$ reflects only some parts of FeSe with observable NMR signals.   

The disappearance of paramagnetic NMR signals below a peak of $1/T_{1}T$ is a typical signature of a magnetic phase transition or spin freezing.  If the Fe magnetic moments are statically ordered at lower temperatures, they would exert well-defined static hyperfine fields $B_{hf} \sim 1.5$~Tesla \cite{Kitagawa} on $^{77}$Se nuclear spins and split the NMR lineshape.  However, we didn't find any additional NMR signals at lower temperatures.  Therefore the disappearance of the NMR signals means that (i) glassy slowing of spin fluctuations makes the longitudinal and transverse relaxation times $T_1$ and $T_2$ of $^{77}$Se NMR signals so fast that spin echo can't form in some parts of FeSe layers, and/or (ii) (nearly) static hyperfine magnetic field $B_{hf}$ has a large distribution.  In the case of the SDW ordered phase in lightly electron-doped BaFe$_{2}$As$_2$, $B_{hf}$ has a continuous distribution up to $\sim 1.5$~Tesla \cite{Ning3}.  In the present context, even if these hyperfine fields are static, the $^{77}$Se NMR linewidth may be as broad as $\gamma_{n}/2\pi \times B_{hf} \sim 12$~MHz, {\it i.e.} the NMR line may be broadened by a factor of $\sim 400$.   In any case, these NMR anomalies above $T_c$ strongly suggest that {\it applied pressure above $\sim 1.4$~GPa enhances spin fluctuations so strongly that a glassy spin freezing takes place in the FeSe layers before bulk superconductivity sets in}.  We recall that high $T_c$ cuprate and URu$_2$Si$_2$ superconductors exhibit analogous situation in the vicinity of  the stripe phase and the hidden ordered phase, respectively \cite{Hunt,Kohori}.  In passing, the inhomogneous electronic properties may be the underlying reason why the superconducting transition in 2.2~GPa becomes  broad, as shown in Fig.1a.  The results in Fig.2d show that the loss of NMR signal intensity may be also present  in 0 and 0.7~GPa somewhat above $T_c$.  However, earlier $\mu$SR measurements  didn't reveal any static magnetic order in a superconducting specimen of "FeSe$_{0.85}$" \cite{Klauss}. The rather abrupt loss of the NMR signal below $T_c$ at 0 and 0.7~GPa may merely be due to the Meissner effect, which limits the NMR intensity by shielding the R.F. pulses for NMR measurements.  

To summarize, we have demonstrated that the electronic properties of the stoichiometric 
FeSe superconductor are very similar to those of optimally electron-doped Ba[Fe$_{0.92}$Co$_{0.08}$]$_{2}$As$_2$ high $T_c$ superconductor.   Contrary to an earlier NMR report on a disordered "FeSe$_{0.92}$" sample \cite{Tou}, our results for superconducting $\beta$-FeSe show no evidence for canonical Fermi liquid above $T_c$, {\it i.e.} the Korringa relation $1/T_{1}TK^{2} = const.$ is not satisfied.  Instead, large enhancement of $1/T_{1}T$ below 100~K  indicates that antiferromagnetic spin fluctuations are strongly enhanced toward $T_c$ in FeSe.  Application of pressure further enhances both spin fluctuations and $T_c$, pointing toward a positive link between antiferromagnetic spin fluctuations and the superconducting mechanism.      

The work at McMaster was supported by NSERC 
and CIFAR.  The work at Princeton was supported primarily by the U.S. Department of Energy, Division of Basic Energy Sciences, Grant DE-FG02-98ER45706, and in part by the NSF-MRSEC program, grant DMR-0819860.\\


\begin{thebibliography}{26}
\expandafter\ifx\csname natexlab\endcsname\relax\def\natexlab#1{#1}\fi
\expandafter\ifx\csname bibnamefont\endcsname\relax
  \def\bibnamefont#1{#1}\fi
\expandafter\ifx\csname bibfnamefont\endcsname\relax
  \def\bibfnamefont#1{#1}\fi
\expandafter\ifx\csname citenamefont\endcsname\relax
  \def\citenamefont#1{#1}\fi
\expandafter\ifx\csname url\endcsname\relax
  \def\url#1{\texttt{#1}}\fi
\expandafter\ifx\csname urlprefix\endcsname\relax\def\urlprefix{URL }\fi
\providecommand{\bibinfo}[2]{#2}
\providecommand{\eprint}[2][]{\url{#2}}

\bibitem[{\citenamefont{Kamihara et~al.}(2008)\citenamefont{Kamihara, Watanabe,
  Hirano, and Hosono}}]{Kamihara}
\bibinfo{author}{\bibfnamefont{Y.}~\bibnamefont{Kamihara}}
  \bibinfo{author}{\bibfnamefont{T.}~\bibnamefont{Watanabe}},
  \bibinfo{author}{\bibfnamefont{M.}~\bibnamefont{Hirano}}, \bibnamefont{and}
  \bibinfo{author}{\bibfnamefont{H.}~\bibnamefont{Hosono}},
  \bibinfo{journal}{J. Amer. Chem. Soc.} \textbf{\bibinfo{volume}{130}},
  \bibinfo{pages}{3296} (\bibinfo{year}{2008}).

\bibitem[{\citenamefont{Norman}(2008)}]{Norman}
\bibinfo{author}{\bibfnamefont{M.}~\bibnamefont{Norman}},
  \bibinfo{journal}{Physics} \textbf{\bibinfo{volume}{1}}, \bibinfo{pages}{21}
  (\bibinfo{year}{2008}).

\bibitem[{\citenamefont{de~la Cruz et~al.}(2008)\citenamefont{de~la Cruz,
  Huang, Lynn, Li, Ratcliff, Zarestky, Mook, Chen, Luo, Wang
  et~al.}}]{delaCruz}
\bibinfo{author}{\bibfnamefont{C.}~\bibnamefont{de~la Cruz}}
  \bibnamefont{et~al.}, \bibinfo{journal}{Nature}
  \textbf{\bibinfo{volume}{453}}, \bibinfo{pages}{899} (\bibinfo{year}{2008}).

\bibitem[{\citenamefont{Rotter et~al.}(2008)\citenamefont{Rotter, Tegel,
  Johrendt, Schellenberg, Hermes, and Pottgen}}]{Rotter}
\bibinfo{author}{\bibfnamefont{M.}~\bibnamefont{Rotter}}
\bibnamefont{et~al.},
  \bibinfo{journal}{Phys. Rev. B} \textbf{\bibinfo{volume}{78}},
  \bibinfo{pages}{020503(R)} (\bibinfo{year}{2008}).

\bibitem[{\citenamefont{Sefat et~al.}(2008)\citenamefont{Sefat, Jin, McGuire,
  Sales, Singh, and Mandrus}}]{Sefat}
\bibinfo{author}{\bibfnamefont{A.~S.} \bibnamefont{Sefat}}
\bibnamefont{et~al.},
  \bibinfo{journal}{Phys. Rev. Lett.} \textbf{\bibinfo{volume}{101}},
  \bibinfo{pages}{117004} (\bibinfo{year}{2008}).

\bibitem[{\citenamefont{Hsu et~al.}(2008)\citenamefont{Hsu, Luo, Yeh, Chen,
  Huang, Wu, Lee, Huang, Chu, Yan et~al.}}]{Hsu}
\bibinfo{author}{\bibfnamefont{F.~C.} \bibnamefont{Hsu}}
  \bibnamefont{et~al.}, \bibinfo{journal}{Proc. Nat. Acad. Sci.}
  \textbf{\bibinfo{volume}{105}}, \bibinfo{pages}{14262}
  (\bibinfo{year}{2008}).

\bibitem[{\citenamefont{McQueen et~al.}(2009)\citenamefont{McQueen, Huang,
  Ksenofotov, Felser, Xu, Zandbergen, Hor, Allred, Williams, Qu
  et~al.}}]{McQueen}
\bibinfo{author}{\bibfnamefont{T.~M.} \bibnamefont{McQueen}}
\bibnamefont{et~al.},
  \bibinfo{journal}{Phys. Rev. B} \textbf{\bibinfo{volume}{79}},
  \bibinfo{pages}{014522} (\bibinfo{year}{2009}).

\bibitem[{\citenamefont{Johannes}(2008)}]{Johannes}
\bibinfo{author}{\bibfnamefont{M.}~\bibnamefont{Johannes}},
  \bibinfo{journal}{Physics} \textbf{\bibinfo{volume}{1}}, \bibinfo{pages}{28}
  (\bibinfo{year}{2008}).

\bibitem[{\citenamefont{Mizuguchi et~al.}(2008)\citenamefont{Mizuguchi,
  Tomioka, Tsuda, Yamaguchi, and Takano}}]{Mizuguchi}
\bibinfo{author}{\bibfnamefont{Y.}~\bibnamefont{Mizuguchi}}
\bibnamefont{et~al.},
  \bibinfo{journal}{Appl. Phys. Lett} \textbf{\bibinfo{volume}{93}},
  \bibinfo{pages}{152505} (\bibinfo{year}{2008}).

\bibitem[{\citenamefont{Medvedev et~al.}()\citenamefont{Medvedev, McQueen,
  Trojan, Palasyuk, Eremets, Cava, Naghavi, Casper, Ksenofontov, Wortmann
  et~al.}}]{Medvedev}
\bibinfo{author}{\bibfnamefont{S.}~\bibnamefont{Medvedev}}
  \bibnamefont{et~al.}, \eprint{cond-mat/0903.2143}.

\bibitem[{\citenamefont{Margadonna et~al.}()\citenamefont{Margadonna,
  Takabayashi, Ohishi, Mizuguchi, Takano, Kagayama, Nakagawa, Takata, and
  Prassides}}]{Margadonna}
\bibinfo{author}{\bibfnamefont{S.}~\bibnamefont{Margadonna}}
  \bibnamefont{et~al.},   \eprint{cond-mat/0903.2204}.

\bibitem[{\citenamefont{Subedi et~al.}(2008)\citenamefont{Subedi, Zhang, Singh,
  and Du}}]{Subedi}
\bibinfo{author}{\bibfnamefont{A.}~\bibnamefont{Subedi}}
\bibnamefont{et~al.},
  \bibinfo{journal}{Phys. Rev. B} \textbf{\bibinfo{volume}{78}},
  \bibinfo{pages}{134514} (\bibinfo{year}{2008}).

\bibitem[{\citenamefont{Kotegawa et~al.}(2008)\citenamefont{Kotegawa, Masaki,
  Awai, Tou, Mizuguchi, and Takano}}]{Tou}
\bibinfo{author}{\bibfnamefont{H.}~\bibnamefont{Kotegawa}}
\bibnamefont{et~al.},
  \bibinfo{journal}{J. Phys. Soc. Jpn.} \textbf{\bibinfo{volume}{77}},
  \bibinfo{pages}{113703} (\bibinfo{year}{2008}).

\bibitem[{\citenamefont{Ahilan et~al.}(2008)\citenamefont{Ahilan, Ning, Imai,
  Sefat, Jin, McGuire, Sales, and Mandrus}}]{Ahilan}
\bibinfo{author}{\bibfnamefont{K.}~\bibnamefont{Ahilan}}
\bibnamefont{et~al.},
  \bibinfo{journal}{Phys. Rev. B} \textbf{\bibinfo{volume}{78}},
  \bibinfo{pages}{100501(R)} (\bibinfo{year}{2008}).

\bibitem[{\citenamefont{Grafe et~al.}()\citenamefont{Grafe, Lang, Hammerath,
  Paar, Manthey, Koch, Rosner, Curro, G.~Behr, Leps et~al.}}]{NJP}
\bibinfo{author}{\bibfnamefont{H.-J.} \bibnamefont{Grafe}}
\bibnamefont{et~al.},
  \bibinfo{journal}{New J. Phys.} \textbf{\bibinfo{volume}{11}},
  \bibinfo{pages}{035002} (\bibinfo{year}{2009}).

\bibitem[{\citenamefont{Ning et~al.}(2008)\citenamefont{Ning, Ahilan, Imai,
  Sefat, Jin, McGuire, Sales, and Mandrus}}]{Ning1}
\bibinfo{author}{\bibfnamefont{F.~L.} \bibnamefont{Ning}}
\bibnamefont{et~al.},
  \bibinfo{journal}{J. Phys. Soc. Jpn.} \textbf{\bibinfo{volume}{77}},
  \bibinfo{pages}{103705} (\bibinfo{year}{2008}).

\bibitem[{\citenamefont{Ning et~al.}(2009)\citenamefont{Ning, Ahilan, Imai,
  Sefat, Jin, McGuire, Sales, and Mandrus}}]{Ning2}
\bibinfo{author}{\bibfnamefont{F.~L.} \bibnamefont{Ning}}
\bibnamefont{et~al.},
  \bibinfo{journal}{J. Phys. Soc. Jpn.} \textbf{\bibinfo{volume}{78}},
  \bibinfo{pages}{013711} (\bibinfo{year}{2009}).

\bibitem[{\citenamefont{Kitagawa et~al.}(2008)\citenamefont{Kitagawa, Katayama,
  Ohgushi, Yoshida, and Takigawa}}]{Kitagawa}
\bibinfo{author}{\bibfnamefont{K.}~\bibnamefont{Kitagawa}}
\bibnamefont{et~al.},
  \bibinfo{journal}{J. Phys. Soc. Jpn.} \textbf{\bibinfo{volume}{77}},
  \bibinfo{pages}{114709} (\bibinfo{year}{2008}).

\bibitem[{\citenamefont{Johnston}(1989)}]{Johnston}
\bibinfo{author}{\bibfnamefont{D.~C.} \bibnamefont{Johnston}},
  \bibinfo{journal}{Phys. Rev. Lett.} \textbf{\bibinfo{volume}{62}},
  \bibinfo{pages}{957} (\bibinfo{year}{1989}).

\bibitem[{\citenamefont{Graser et~al.}()\citenamefont{Graser, Maier,
  Hirschfeld, and Scalapino}}]{Scalapino}
\bibinfo{author}{\bibfnamefont{S.}~\bibnamefont{Graser}}
\bibnamefont{et~al.},
  \eprint{cond-mat/0812.0343}.

\bibitem[{\citenamefont{Shastry}(1989)}]{Shastry}
\bibinfo{author}{\bibfnamefont{B.~S.} \bibnamefont{Shastry}},
  \bibinfo{journal}{Phys. Rev. Lett.} \textbf{\bibinfo{volume}{63}},
  \bibinfo{pages}{1288} (\bibinfo{year}{1989}).

\bibitem[{\citenamefont{Nakai et~al.}(2008)\citenamefont{Nakai, Ishida,
  Kamihara, Hirano, and Hosono}}]{Nakai}
\bibinfo{author}{\bibfnamefont{Y.}~\bibnamefont{Nakai}}
\bibnamefont{et~al.},
  \bibinfo{journal}{J. Phys. Soc. Jpn.} \textbf{\bibinfo{volume}{77}},
  \bibinfo{pages}{073701} (\bibinfo{year}{2008}).

\bibitem[{\citenamefont{Ning et~al.}()\citenamefont{Ning, Ahilan, Imai, Sefat,
  Jin, McGuire, Sales, and Mandrus}}]{Ning3}
\bibinfo{author}{\bibfnamefont{F.~L.} \bibnamefont{Ning}}
\bibnamefont{et~al.},
  \eprint{cond-mat/0902.1788}.

\bibitem[{\citenamefont{Hunt et~al.}(2001)\citenamefont{Hunt, Singer,
  Cederstrom, and Imai}}]{Hunt}
\bibinfo{author}{\bibfnamefont{A.~W.} \bibnamefont{Hunt}}
 \bibnamefont{et~al.},
  \bibinfo{journal}{Phys. Rev. B} \textbf{\bibinfo{volume}{64}},
  \bibinfo{pages}{134525} (\bibinfo{year}{2001}).

\bibitem[{\citenamefont{Matsuda et~al.}(2001)\citenamefont{Matsuda, Kohori,
  Kohara, Kuwahara, and Amitsuka}}]{Kohori}
\bibinfo{author}{\bibfnamefont{K.}~\bibnamefont{Matsuda}}
\bibnamefont{et~al.},
  \bibinfo{journal}{Phys. Rev. Lett.} \textbf{\bibinfo{volume}{87}},
  \bibinfo{pages}{087203} (\bibinfo{year}{2001}).

\bibitem[{\citenamefont{Khasanov et~al.}()\citenamefont{Khasanov, Conder,
  Pomjakushina, Amato, Baines, Bukowski, Karpinski, Katrych, Klauss, Luetkens
  et~al.}}]{Klauss}
\bibinfo{author}{\bibfnamefont{R.}~\bibnamefont{Khasanov}}
  \bibnamefont{et~al.}, \eprint{cond-mat/0810.1716}.

\end{thebibliography}

\end{document}